\documentclass{article}
\usepackage[utf8]{inputenc}
\usepackage{csquotes}

\usepackage{amssymb}
\usepackage{amsmath}
\usepackage{color}
\usepackage{amsthm}
\theoremstyle{definition} 
\theoremstyle{plain} \newtheorem{proposition}{Proposition}
\theoremstyle{plain} 
\theoremstyle{definition} 
\theoremstyle{remark} 
\theoremstyle{remark} 
\theoremstyle{plain} \newtheorem{lemma}[proposition]{Lemma}
\theoremstyle{plain} \newtheorem{procedure}{Procedure}

\usepackage{geometry}
\usepackage{hyperref}

\usepackage{comment}

\usepackage{graphicx}

\title{Deviations from normality in a financial model without short-selling}
\author{Nahuel I. Arca}

\begin{document}

\sloppy

\maketitle

\begin{abstract}
    We present a variation of the well-known binomial model of asset prices. This variation is based on the interaction of a finite number of investors (grouped in two kinds, bulls and bears) with a single market maker, in an environment with bounds to short selling. We prove a formula for all the moments of the logarithmic returns and from that we derive the corresponding formula for the binomial model. As an application of the model, we show how to compute parameters in order to approximate given moments, enabling the modeling of skewness and excess kurtosis. We provide a concrete example of this procedure, using real data from a stock, and we present two plots revealing its convergence pattern. Finally, we generalize the framework to account for the possibility of more heterogeneous investors, and give the corresponding formula for the moments of the logarithmic returns, and the algorithm for fitting given moments.
\end{abstract}

\section{Introduction}

Usually, one's first approach to the world of mathematical finance is through the binomial model \cite{cox1979option,rendleman1979two,sharpe1999investment}. Such model can be understood as a struggle between two groups of investors: the \emph{bulls} and the \emph{bears}. The bulls are those investors who believe that the asset's price will go up, while the bears are those who believe that the asset's price will go down. Let's suppose that the proportion of bulls between all the investors is $p_u$, while the proportion of bears between all the investors is $p_d(=1-p_u)$. At every instant of time, an investor is randomly chosen:
\begin{itemize}
    \item If a bull was chosen, he/she buys the asset, making the price go up by a factor $u>1$.
    \item If a bear was chosen, she sells the asset, making the price go down by a factor $d<1$.
\end{itemize}
In this way, if $S(t)$ is the asset's price at time $t$, we have that
\begin{align*}
    \mathbb{P}(S(t)=uS(t-1))=p_u\qquad\text{and}\qquad\mathbb{P}(S(t)=dS(t-1))=p_d\text{ .}
\end{align*}
The probabilities $p_u$ and $p_d$ are fixed. A model in line with this interpretation can be found in section 4 of \cite{cont2000herd}.\newline
Under this interpretation, the binomial model implicitly assumes at least one of the following hypotheses:
\begin{itemize}
    \item The amount of investors is sufficiently large: if a bull spends all of her money, or a bear sells all of her stocks, this doesn't significantly alter the values of $p_u$ and $p_d$.
    \item The buying of stocks by the bulls, and the selling of stocks by the bears, is carried out in a gradual manner: it takes time for a bull to spend all of her money, or for a bear to sell all of her stocks.
    \item Every investor can borrow money and stocks without bounds: a bull can keep on buying even if she has no money, and a bear can keep on selling even if she has no stocks.
    \item Continuously, the investors perceive incomes of money and stocks outside the considered market: bulls always have money and bears always have stocks.
\end{itemize}

On the other hand, a model based on the finiteness of the investors' assets, that is to say on the violation of the just mentioned hypotheses, is the one presented in \cite{caginalp2003theoretical}. It has continuous time and different investors' groups. At time $t$, each group has a parameter $k(t)\in (0,1)$ that determines the proportion in which they buy or sell assets based on considerations about fundamental and technical factors, and a magnitude $B(t)\in (0,1)$ that represents the proportion of their wealth invested in assets. These functions determine the supply $S(t)$ and the demand $D(t)$, and the price $P(t)$ changes according to the equation
\begin{align*}
    \frac{d}{dt}\log P(t)=\log\left(\frac{D(t)}{S(t)}\right)\text{ ,}
\end{align*}
which closes a system of deterministic differential equations. In particular, $k(t)$ is given by the equation
\begin{align*}
    k(t)=\frac{1+\tanh\zeta(t)}{2}\text{ ,}
\end{align*}
where $\zeta(t)\in\mathbb{R}$ is a parameter that represents the investor's preference on buying or not stocks, and the equation guarantees that $k(t)\in (0,1)$. $B(t)$ changes according to the equation
\begin{align*}
    \frac{dB}{dt}=k\cdot(1-B)-(1-k)\cdot B+B(1-B)\frac{1}{P}\frac{dP}{dt}\text{ ,}
\end{align*}
which guarantees that starting with $B(0)\in (0,1)$, then $B(t)\in(0,1)$ for all $t$, which means that investors neither short nor borrow. Also, the changes in $B(t)$ are due only to buying and selling of stocks and price changes, which indicates that neither money nor stocks enter the considered market from the outside. The other two hypotheses are neither necessary.\newline 
In this model, when the money of a group of investors becomes scarce, they can't keep on pushing the stock price up and, analogously, when their stocks become scarce, they can't keep on pushing the price down. When two groups of investors have two different fundamental values, the first one larger and the second one smaller than the current price of the asset, a struggle takes place that eventually depletes the money of the first group or the stocks of the second group, and the price ends up converging to the fundamental value of the struggle's winner. This phenomenon is used in \cite{caginalp2003theoretical} to explain some known patterns of technical analysis.\newline 

In this work, an approach is developed, that combines features of both previously mentioned models. In section \ref{palabras}, the model is described as a representation of a real scenario. In such scenario there are two groups of investors, the bulls and the bears. The bulls buy stocks, while the bears sell them. Mediating the transactions between the groups, there is a market maker that fixes the prices. The technical details on the formalization can be found in the appendix.

In the binomial model, the distribution of the logarithmic returns can be approximated with the central limit theorem. A way to measure how much such distributions deviate in this new model is to study their moments\cite{akhiezer2020classical,krein1977markov,schmudgen2017moment,shohat1950problem}. With this in mind, in section \ref{momentos} we prove a formula that allows us to compute all these moments.

Although the distribution of the logarithmic returns in the binomial model can be approximated with the central limit theorem, its moments may also be computed exactly. This is what we do in section \ref{binomial}, by taking limit in a formula introduced in section \ref{momentos}.

As an application of the model, in section \ref{fit_mom}, we show how to take parameter values in order to approximate given moments arbitrarily close. This procedure is exemplified with the use of real financial data, which also suggests a clear convergence pattern.

In section \ref{multigrupos}, we generalize the framework introduced, for the case in which there are several groups of bulls with different amounts of money and/or several groups of bears with different amounts of stocks. The generalization of the formula deduced in section \ref{momentos} and the generalization of the algorithm presented in section \ref{fit_mom} are introduced, observing that analogous proofs hold for this general case.

\section{The model in words}\label{palabras}

We will model the following scenario. There are $N$ investors interacting in a market. Besides the investors, there is a \emph{market maker} (MM) always willing to buy and sell stocks. At time $t$, the investors can be divided in $3$ groups:

\begin{enumerate}
    \item There are $N_u(t)$ \emph{bulls}: these are investors with money that want to buy stocks.
    \item There are $N_d(t)$ \emph{bears}: these are investors with stocks that want to sell them.
    \item There are $N-N_u(t)-N_d(t)$ \emph{inactives}: these are investors without money that would like to buy stocks, and investors without stocks that would like to sell them.
\end{enumerate}

At time $t$ an investor is randomly chosen, and every investor has the same probability of being chosen.

Let $p(t)$ be the stock price at time $t$. The MM operates with two constants $0<d<1<u$. We shall assume that, at time $t$,
\begin{enumerate}
    \item if a bull is chosen, she spends all her money buying stocks from the MM, who sells them to her at price $up(t)$, and this becomes the new stock price, that is $p(t+1)=up(t)$;
    \item if a bear is chosen, she sells all her stocks to the MM, who buys them from her at price $dp(t)$, and this becomes the new stock price, that is $p(t+1)=dp(t)$;
    \item if an inactive investor is chosen, stocks are neither bought nor sold, and the price keeps its current level, that is $p(t+1)=p(t)$.
\end{enumerate}

The chosen investor becomes inactive, whatever was his origin:
\begin{enumerate}
    \item if she was a bull, $N_u(t+1)=N_u(t)-1$ and $N_d(t+1)=N_d(t)$;
    \item if she was a bear, $N_d(t+1)=N_d(t)-1$ and $N_u(t+1)=N_u(t)$;
    \item if she was an inactive investor, $N_u(t+1)=N_u(t)$ and $N_d(t+1)=N_d(t)$.
\end{enumerate}

A formalization of this model is provided in the appendix.

\section{Computation of moments of the logarithmic returns}\label{momentos}

In these kind of models, practitioners are usually interested in the distribution of the logarithmic returns. In this case, $\log(p(t)/p(0))$ is a random variable with compact support (its support is actually finite), therefore its moments determine its distribution: the well-known uniqueness result of the Hausdorff moment problem \cite{feller1991introduction}.

The main goal of this section is to prove a formula that allows us to easily compute the first moments of the logarithmic returns. That formula is
\begin{equation}\label{form_4}
\begin{split}
\mathbb{E}\left(\log\left(\frac{p(t)}{p(0)}\right)^n\right)=\sum_{\substack{n_u,n_d\in\mathbb{N}_0\\n_u+n_d=n}}\frac{n!}{n_u!n_d!}(\log u)^{n_u}(\log d)^{n_d}\sum_{k_u=0}^{n_u}\frac{N_u(0)!}{(N_u(0)-k_u)!}\left\{\begin{matrix}n_u\\ k_u\end{matrix}\right\}\\
\cdot\sum_{k_d=0}^{n_d}\frac{N_d(0)!}{(N_d(0)-k_d)!}\left\{\begin{matrix}n_d\\ k_d\end{matrix}\right\}\sum_{j=0}^{k_u+k_d}(-1)^j\binom{k_u+k_d}{j}\left(1-\frac{j}{N}\right)^{t}\text{ ,}
\end{split}
\end{equation}

where the symbol $\left\{\begin{matrix}n\\ k\end{matrix}\right\}$ represents a Stirling number of the second kind\cite{patashnik1994concrete}. The reader can find Python code for computing these values in \url{https://github.com/nahueliarca/bullsvsbears/tree/main}.\newline
In order to prove this formula, we need some preparatory lemmas, whose proofs can be found in section \ref{app_fit_mom}, for the reader's convenience.\newline

Observe that
\begin{align*}
\log\left(\frac{p(t)}{p(0)}\right)&=\sum_{i=1}^t \log\left(\frac{p(i)}{p(i-1)}\right)\text{ ,}
\end{align*}
and let $X_i:=\log(p(i)/p(i-1))$. We have the following result:

\begin{lemma}\label{lemma_first_formula}
Let $0=i_0<i_1<\ldots<i_k<i_{k+1}$ be integers, then
\begin{equation}\label{prop_3}
\begin{split}
\mathbb{E}\left(\frac{N_u(i_{k+1}-1)!}{(N_u(i_{k+1}-1)-l_u)!}\frac{N_d(i_{k+1}-1)!}{(N_d(i_{k+1}-1)-l_d)!}\prod_{j=1}^{k}X_{i_j}^{n_j}\right)\\
=\frac{1}{N^k}\prod_{j=1}^{k+1}\left(1-\frac{l_u+l_d+k-j+1}{N}\right)^{i_j-i_{j-1}-1}\\
\cdot\sum_{s\in\{u,d\}^k}\frac{N_u(0)!}{(N_u(0)-l_u-\#\{s_j=u\})!}\frac{N_d(0)!}{(N_d(0)-l_d-\#\{s_j=d\})!}\prod_{j=1}^k(\log s_j)^{n_j}\text{ .}
\end{split}
\end{equation}
\end{lemma}

With this result at our disposal, let us give the first steps towards proving formula \eqref{form_4}. We proceed by direct computation of the logarithmic returns. It is straightforward to see that
\begin{align*}
\mathbb{E}\left(\log\left(\frac{p(t)}{p(0)}\right)^n\right)=\sum_{k=1}^n\sum_{\substack{n_i\in\mathbb{N}\\n_1+\ldots+n_k=n}}\sum_{1\leq i_1<\ldots<i_k\leq t}\frac{n!}{n_1!\cdots n_k!}\mathbb{E}\left(\prod_{j=1}^{k}X_{i_j}^{n_j}\right)\text{ .}
\end{align*}
We can rewrite the terms of this sum, using lemma \ref{lemma_first_formula}: taking $l_u=l_d=0$, we get the formula
\begin{align*}
\mathbb{E}\left(\prod_{j=1}^{k}X_{i_j}^{n_j}\right)=\frac{1}{N^k}\prod_{j=1}^{k}\left(1-\frac{k-j+1}{N}\right)^{i_j-i_{j-1}-1}\\
\cdot\sum_{s\in\{u,d\}^k}\frac{N_u(0)!}{(N_u(0)-\#\{s_j=u\})!}\frac{N_d(0)!}{(N_d(0)-\#\{s_j=d\})!}\prod_{j=1}^k(\log s_j)^{n_j}\text{ ,}
\end{align*}
hence
\begin{align*}
\mathbb{E}\left(\log\left(\frac{p(t)}{p(0)}\right)^n\right)=\sum_{k=1}^n\sum_{\substack{n_i\in\mathbb{N}\\n_1+\ldots+n_k=n}}\sum_{s\in\{u,d\}^k}\sum_{\substack{m_i\in\mathbb{N}\\m_1+\ldots+m_k\leq t}}\frac{n!}{n_1!\cdots n_k!}\frac{1}{N^k}\\
\cdot\prod_{j=1}^k\left(1-\frac{k-j+1}{N}\right)^{m_j-1}\frac{N_u(0)!}{(N_u(0)-\#\{s_j=u\})!}\frac{N_d(0)!}{(N_d(0)-\#\{s_j=d\})!}\prod_{j=1}^k(\log s_j)^{n_j}\text{ .}
\end{align*}

For practical applications, we need an expression easy to compute for large values of $t$ and small values of $n$. This last expression is hard to compute for large values of $t$, so this section ends with a rewriting of this formula. A first step in that direction is provided by the following result.

\begin{lemma}\label{coro_mom_tri}
\begin{align}
\mathbb{E}\left(\log\left(\frac{p(t)}{p(0)}\right)^n\right)=\sum_{\substack{n_u,n_d\in\mathbb{N}_0\\n_u+n_d=n}}(\log u)^{n_u}(\log d)^{n_d}\sum_{\substack{n_u^i\in\mathbb{N}\\n_u^1+\ldots+n_u^{k_u}=n_u}}\frac{N_u(0)!}{N^{k_u}(N_u(0)-k_u)!}\nonumber\\
\sum_{\substack{n_d^i\in\mathbb{N}\\n_d^1+\ldots+n_d^{k_d}=n_d}}\frac{N_d(0)!}{N^{k_d}(N_d(0)-k_d)!}\frac{(k_u+k_d)!}{k_u!k_d!}\frac{n!}{n_u^1!\cdots n_u^{k_u}!\cdot n_d^1!\cdots n_d^{k_d}!}\nonumber\\
\sum_{\substack{m_i\in\mathbb{N}_0\\m_0+\ldots+m_{k_u+k_d}=t-k_u-k_d}}\prod_{j=0}^{k_u+k_d}\left(1-\frac{j}{N}\right)^{m_j}\text{ .}\label{mom1}
\end{align}
\end{lemma}

In order to make this formula easy to compute, we would like to get rid of the last three sums. The following lemma helps us deal with the last one.

\begin{lemma}\label{comb_lemma}
Given $k\geq 0$, and $c_0,\ldots,c_k\in\mathbb{R}$ pairwise different, then
\begin{align*}
\sum_{m_0+\ldots+m_k=m}\prod_{j=0}^{k}c_j^{m_j}=\sum_{j=0}^k\frac{c_j^{m+k}}{\prod_{i\neq j}(c_j-c_i)}\text{ .}
\end{align*}
\end{lemma}

Applying this lemma on \eqref{mom1}, we get
\begin{equation}\label{form_4_1}
\begin{split}
\mathbb{E}\left(\log\left(\frac{p(t)}{p(0)}\right)^n\right)=\sum_{\substack{n_u,n_d\in\mathbb{N}_0\\n_u+n_d=n}}(\log u)^{n_u}(\log d)^{n_d}\sum_{\substack{n_u^i\in\mathbb{N}\\n_u^1+\ldots+n_u^{k_u}=n_u}}\binom{N_u(0)}{k_u}\\
\cdot\sum_{\substack{n_d^i\in\mathbb{N}\\n_d^1+\ldots+n_d^{k_d}=n_d}}\binom{N_d(0)}{k_d}\frac{n!}{n_u^1!\cdots n_u^{k_u}!\cdot n_d^1!\cdots n_d^{k_d}!}\sum_{j=0}^{k_u+k_d}(-1)^j\binom{k_u+k_d}{j}\left(1-\frac{j}{N}\right)^{t}\text{ .}
\end{split}
\end{equation}

This expression can be simplified with the use of the Stirling numbers of the second kind. Recall that $\left\{\begin{matrix}n\\ k\end{matrix}\right\}$ is the number of partitions of a set of $n$ elements in $k$ non-empty sets. Hence, the number of surjective functions from a set of $n$ elements to a set of $k$ elements is $k!\left\{\begin{matrix}n\\ k\end{matrix}\right\}$. Calling the cardinal of the preimage of the codomain's $i$-th element $n_i$, we get
\begin{align*}
k!\left\{\begin{matrix}n\\ k\end{matrix}\right\}=\sum_{\substack{n_i\in\mathbb{N}\\ n_1+\ldots+n_k=n}}\frac{n!}{n_1!\cdots n_k!}\text{ .}
\end{align*}
Using this identity on \eqref{form_4_1} yields \eqref{form_4}.

\section{Moments of the model with infinitely many investors}\label{binomial}

As a limit case of our framework, the binomial model can be recovered. Indeed,
\begin{align*}
    \mathbb{P}\left(X_i=\log u|N_u(i-1)=x\right)=\frac{x}{N}\qquad\text{and}\\
    N_u(0)\geq N_u(i-1)\geq N_u(0)-i+1\text{ ,}
\end{align*}
so if we take $N_u(0)=q_u N$ for a given $q_u\in (0,1)$, then we get
\begin{align*}
    \mathbb{P}\left(X_i=\log u|N_u(i-1)=x\right)\to q_u
\end{align*}
as $N\to\infty$. An analogous definition for $N_d(0)$ yields the existence of limits for $\mathbb{P}\left(X_i=\log d|N_d(i-1)=y\right)$ and $\mathbb{P}\left(X_i=0|N_u(i-1)=x, N_d(i-1)=y\right)$.

Taking advantage of this, in this section we compute the logarithmic returns of the binomial model. These results do not appear in the current literature, to the best of our knowledge.

Let $q_u$ and $q_d$ be nonnegative numbers such that $q_u+q_d\leq 1$. Consider $N_u(0):=q_uN$ and $N_d(0):=q_dN$. Then from equation \eqref{mom1}, $\mathbb{E}\left(\log\left(p(t)/p(0)\right)^n\right)$ converges to
\begin{align*}
    \sum_{\substack{n_u,n_d\in\mathbb{N}_0\\n_u+n_d=n}}\frac{n!}{n_u!n_d!}(\log u)^{n_u}(\log d)^{n_d}\sum_{k_u=0}^{n_u}q_u^{k_u}\left\{\begin{matrix}n_u\\ k_u\end{matrix}\right\}\sum_{k_d=0}^{n_d}q_d^{k_d}\left\{\begin{matrix}n_d\\ k_d\end{matrix}\right\}\frac{t!}{(t-k_u-k_d)!}\text{ ,}
\end{align*}
as $N\to\infty$. In particular, in the binomial model we have $q_d=1-q_u$, so the $n$-th moment of the logarithmic return at time $t$ is
\begin{align*}
    \sum_{\substack{n_u,n_d\in\mathbb{N}_0\\n_u+n_d=n}}\frac{n!}{n_u!n_d!}(\log u)^{n_u}(\log d)^{n_d}\sum_{k_u=0}^{n_u}q_u^{k_u}\left\{\begin{matrix}n_u\\ k_u\end{matrix}\right\}\sum_{k_d=0}^{n_d}(1-q_u)^{k_d}\left\{\begin{matrix}n_d\\ k_d\end{matrix}\right\}\frac{t!}{(t-k_u-k_d)!}\text{ .}
\end{align*}

\section{Fitting moments}\label{fit_mom}

Part of the importance of financial models for practitioners lie in better understanding financial phenomena and better fitting empirical data. In this section we show how to fit empirical data, and we hope that the final result may shine some light in a possible explanation for known deviations from normality, grounded on real world features such as finitude of assets and limitations to short-selling.

Given the first moments of a distribution, we would like to find parameters $\log u,\log d,N_u(0),N_d(0),N$ and $t$ such that the moments of $\log(p(t)/p(0))$ fit the given values.\newline

Because the framework has 6 parameters, we suspect that it would be possible to fit the first 6 moments, but we didn't find an analytical way to do this. Instead, we rely on a limit taking $t,N\to\infty$ in a particular fashion and show the procedure to fit the first 4 moments.

Assume that we are given the first $4$ sample moments of the logarithmic returns: $m_n$ for $1\leq n\leq 4$. Let
\begin{align*}
	v_1(x)&:=1+\sum_{k=1}^{4}\frac{m_k}{k!}x^k\text{ ,}\\
	u_1(x)&:=\log v_1(x)\qquad\text{and}\\
	\kappa_n&:=u_1^{(n)}(0)\text{ .}
\end{align*}
Let $q(x)$ be the polynomial
\begin{align*}
	q(x):=\det\left(\begin{pmatrix}
    	\kappa_1&\kappa_2\\
    	\kappa_2&\kappa_3
	\end{pmatrix}x-\begin{pmatrix}
    	\kappa_2&\kappa_3\\
    	\kappa_3&\kappa_4
	\end{pmatrix}\right)\text{ ,}
\end{align*}
and let $r_1$ and $r_2$ be its roots. Let
\begin{align*}
	V&:=\begin{pmatrix}
    	1&r_1\\
    	1&r_2
	\end{pmatrix}\text{ ,}\\
	D&:=\begin{pmatrix}
    	r_1&0\\
    	0&r_2
	\end{pmatrix}\qquad\text{and}\\
	\lambda&:=D^{-1}(V^T)^{-1}\begin{pmatrix}
    	\kappa_1\\
    	\kappa_2
	\end{pmatrix}\text{ .}
\end{align*}

Under the following conditions, we can approximate the given sample moments:
\begin{itemize}
	\item $r_1,r_2\in\mathbb{R}\backslash\{0\}$,
	\item $r_1\neq r_2$ and
	\item $\lambda_1,\lambda_2>0$.
\end{itemize}

Assume these conditions hold, we show how to approximate the given moments. Assume w.l.o.g. that $r_1<r_2$.

Let us say we want to approximate the first 4 moments with an error less than some given $\varepsilon>0$.

\begin{procedure}\label{proc_fit_mom}
    \begin{enumerate}
        \item Set $C\gg 0$.\label{step_Nu}
        \item Set $N_d(0)=\lfloor\lambda_1 C\rfloor$, $N_u(0)=\lfloor\lambda_2 C\rfloor$, $d=\exp r_1$ and $u=\exp r_2$.
        \item Compute the first 4 moments using the expression
        \begin{align*}
            \sum_{\substack{n_u,n_d\in\mathbb{N}_0\\n_u+n_d=n}}\frac{n!}{n_u!n_d!}r_1^{n_d}r_2^{n_u}\sum_{k_u=0}^{n_u}\frac{N_u(0)!}{(N_u(0)-k_u)!}\left\{\begin{matrix}n_u\\ k_u\end{matrix}\right\}\sum_{k_d=0}^{n_d}\frac{N_d(0)!}{(N_d(0)-k_d)!}\left\{\begin{matrix}n_d\\ k_d\end{matrix}\right\}C^{-(k_u+k_d)}\text{ .}
        \end{align*}
        If the computed moments don't approximate the sample moments with an error less than $\varepsilon$, go back to step \ref{step_Nu} and take a larger $C$.
        \item Set $N\geq N_u(0)+N_d(0)$.\label{step_N}
        \item Set $t=\lfloor-N\log\left(1-C^{-1}\right)\rfloor$.
        \item Compute the first 4 moments using equation \eqref{form_4}. If the computed moments don't approximate the sample moments with an error less than $\varepsilon$, go back to step \ref{step_N} and take a larger $N$.
    \end{enumerate}
\end{procedure}

To prove the correctness of this procedure, first observe that keeping $C$ fixed, $\mathbb{E}\left(\log\left(p(t)/p(0)\right)^n\right)$ converges to
\begin{align*}
	\sum_{\substack{n_u,n_d\in\mathbb{N}_0\\n_u+n_d=n}}\frac{n!}{n_u!n_d!}r_1^{n_d}r_2^{n_u}\sum_{k_u=0}^{n_u}\frac{N_u(0)!}{(N_u(0)-k_u)!}\left\{\begin{matrix}n_u\\ k_u\end{matrix}\right\}\sum_{k_d=0}^{n_d}\frac{N_d(0)!}{(N_d(0)-k_d)!}\left\{\begin{matrix}n_d\\ k_d\end{matrix}\right\}C^{-(k_u+k_d)}
\end{align*}
as $N\to\infty$. In turn, this last expression converges to
\begin{align*}
	\sum_{\substack{n_{1},n_{2}\in\mathbb{N}_0\\n_{1}+n_{2}=n}}\frac{n!}{n_{1}!n_{2}!}r_1^{n_{1}}r_2^{n_{2}}\sum_{k_1=0}^{n_{1}}\left\{\begin{matrix}n_{1}\\ k_1\end{matrix}\right\}\lambda_1^{k_1}\sum_{k_2=0}^{n_{2}}\left\{\begin{matrix}n_{2}\\ k_2\end{matrix}\right\}\lambda_2^{k_2}
\end{align*}
as $C\to\infty$. Let $X_1$ and $X_2$ be independent random variables following the Poisson distribution with parameters $\lambda_1$ and $\lambda_2$ respectively. Then, according to \cite{riordan1937moment},

\begin{align*}
	\mathbb{E}(X_h^{n_h})&=\sum_{k_h=0}^{n_h}\left\{\begin{matrix}n_h\\ k_h\end{matrix}\right\}\lambda_h^{k_h}\text{ .}
\end{align*}
So, setting
\begin{align*}
	Y:=r_1 X_1+r_2 X_2\,\text{,}
\end{align*}
we get
\begin{align*}
	\mathbb{E}\left(\log\left(\frac{p(t)}{p(0)}\right)^n\right)\to\sum_{\substack{n_{1},n_{2}\in\mathbb{N}_0\\n_{1}+n_{2}=n}}\frac{n!}{n_{1}!n_{2}!}r_1^{n_{1}}r_2^{n_{2}}\mathbb{E}(X_1^{n_1})\mathbb{E}(X_2^{n_2})=\mathbb{E}(Y^{n})
\end{align*}
as $N\to\infty$.

To complete the proof of correctness of procedure \ref{proc_fit_mom}, we have the following result:

\begin{proposition}\label{prop_mom}
For $1\leq n\leq 4$,
	\begin{align*}
    	\mathbb{E}(Y^{n})=m_n\text{ .}
	\end{align*}
\end{proposition}

The proof of this proposition can be found in section \ref{app_fit_mom}, for the reader's convenience.

\subsection{Test with real data}\label{test_real_data}

In this section we apply procedure~\ref{proc_fit_mom} to a real case. As pointed out in \cite{cont2000herd}, fat tails can be observed in intraday returns, as longer time intervals seem more likely to allow for the aggregation of independent identically distributed variables, leading to normally distributed returns. Hence, we analyzed daily return data.

We analyzed the data of three Big Tech companies and an Argentine e-commerce company. At first, we selected the Argentine company, as we thought that it would be more vulnerable to short-selling limitations. In this case, we found that the roots of polynomial $q$ were non-real and decided to try our hand with some well-known companies.

We tried the procedure with the tickers MELI, GOOG, AAPL and MSFT between the dates 2023-09-01 and 2023-10-30. For each ticker, we computed the daily logarithmic returns, those obtained starting with the opening price and finishing with the closing price. From that, we computed the sample moments. In table~\ref{tab:skew_kurt} we can appreciate the skewness and excess kurtosis of the selected stocks' logarithmic returns. The values of the table are clearly non-zero, which underscores their non-normal distribution.
\begin{table}[h!]
    \centering
    \begin{tabular}{l|c|c}
        Ticker & Skewness & Excess kurtosis \\ \hline
        MELI & 0.3837 & -0.5337\\
        GOOG & -0.1467 & -0.3921\\
        AAPL & -0.6373 & -0.2251\\
        MSFT & -0.2997 & 0.8309
    \end{tabular}
    \caption{Skewness and excess kurtosis of the daily logarithmic returns between the dates 2023-09-01 and 2023-10-30}
    \label{tab:skew_kurt}
\end{table}

With all four tickers, we went on with the procedure. With the exception of MSFT, the roots of polynomial $q$ turned up being non-real. Hence, we applied the procedure only for MSFT.

We wanted to get a model with less than 1\% error with respect to the computed sample moments. Values of $C = 10^3$ and $N = 10^6$ were enough for this. The corresponding model parameters are $d = 0.986529\text{, }u=1.01025\text{, }N_d(0)=596\text{, }N_u(0)=704\text{ and }t=1001$. In table \ref{tab:MSFT} we present the moments of the sample, the fit and their relative difference.
\begin{table}[h!]
    \centering
    \begin{tabular}{c|c|c|c}
        Moment order & Sample & Fit & Relative difference\\ \hline
         1 & $-9.051\cdot 10^{-4}$ & $-9.064\cdot 10^{-4}$ & $0.1494\%$\\
         2 & $1.837\cdot 10^{-4}$ & $1.835\cdot 10^{-4}$ & $-0.06762\%$\\
         3 & $-1.238\cdot 10^{-6}$ & $-1.236\cdot 10^{-6}$ & $-0.1564\%$\\
         4 & $1.316\cdot 10^{-7}$ & $1.312\cdot 10^{-7}$ & $-0.3255\%$
    \end{tabular}
    \caption{Parameter fit to the daily logarithmic returns of MSFT between the dates 2023-09-01 and 2023-10-30}
    \label{tab:MSFT}
\end{table}

Let's proceed to analyze the convergence. Procedure~\ref{proc_fit_mom} consists of two parts: steps 1 to 3 consist of finding a large value of $C$ so that
\begin{align*}
    \mu(C,n):=\sum_{\substack{n_u,n_d\in\mathbb{N}_0\\n_u+n_d=n}}\frac{n!}{n_u!n_d!}r_1^{n_d}r_2^{n_u}\sum_{k_u=0}^{n_u}\frac{N_u(0)!}{(N_u(0)-k_u)!}\left\{\begin{matrix}n_u\\ k_u\end{matrix}\right\}\sum_{k_d=0}^{n_d}\frac{N_d(0)!}{(N_d(0)-k_d)!}\left\{\begin{matrix}n_d\\ k_d\end{matrix}\right\}C^{-(k_u+k_d)}
\end{align*}
differs little from $m_n$, and steps 4 to 6 consist of finding a large value of $N$ so that $\mathbb{E}(\log(p(t)/p(0))^n)$ differs little from $\mu(C,n)$ (and correspondingly from $m_n$).

\begin{figure}[h]
    \centering
    \includegraphics[width=1\linewidth]{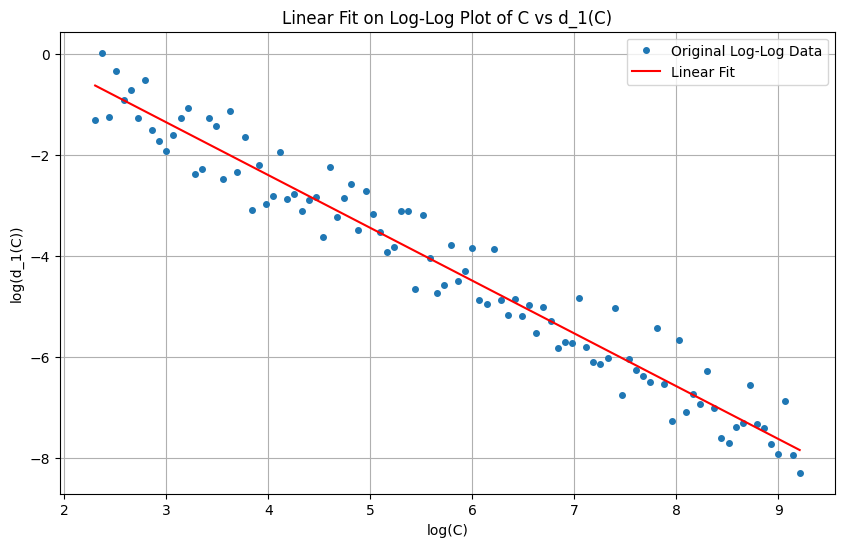}
    \caption{$\log(d_1(C))\sim -1.044\log(C)+1.774$}
    \label{fig:conver_C}
\end{figure}

We define
\begin{align*}
    d_1(C):=\max_{1\leq n\leq 4}\left|\frac{\mu(C,n)-m_n}{m_n}\right|\,\text{,}
\end{align*}
the maximum of the relative differences between $\mu(C,n)$ and $m_n$. Taking different values of $C$ and computing $d_1(C)$ for the MSFT data, we get the plot in figure~\ref{fig:conver_C}.

The data seem to fit the model
\begin{align*}
    \log(d_1(C))\sim -1.044\log(C)+1.774
\end{align*}
with $R^2=0.952$.

\begin{figure}[h]
    \centering
    \includegraphics[width=1\linewidth]{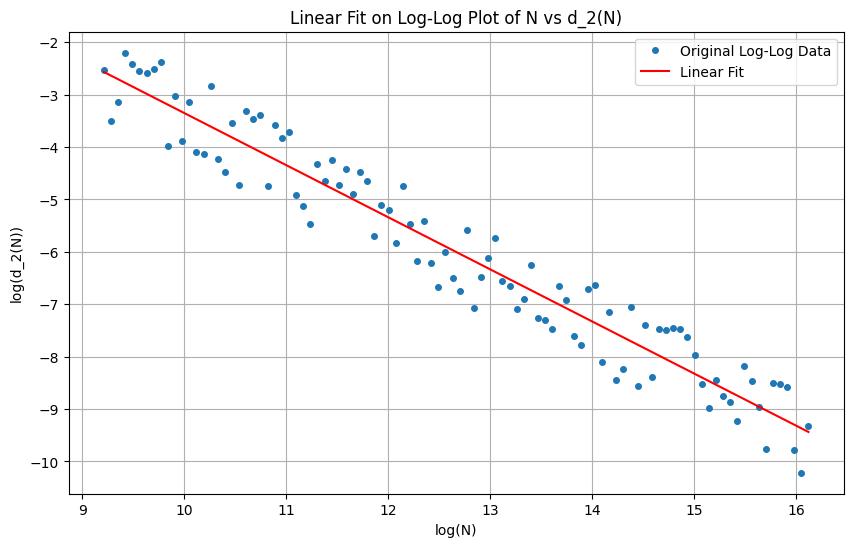}
    \caption{$\log(d_2(N))\sim -0.9946\log(N)+6.594$}
    \label{fig:conver_N}
\end{figure}

Fixing $C=10^3$, we define
\begin{align*}
    d_2(N):=\max_{1\leq n\leq 4}\left|\frac{\mathbb{E}(\log(p(t)/p(0))^n)-\mu(C,n)}{\mu(C,n)}\right|\,\text{,}
\end{align*}
the maximum of the relative differences between $\mathbb{E}(\log(p(t)/p(0))^n)$ and $\mu(C,n)$. Doing the same plot for $N$ and $d_2(N)$, we get figure~\ref{fig:conver_N}.

The data seem to fit the model
\begin{align*}
    \log(d_2(N))\sim -0.9946\log(N)+6.594
\end{align*}
with $R^2=0.9336$.

\section{Multi-group model}\label{multigrupos}

In this section, we present a generalization of the framework.\newline
Assume that there are several groups of bears and bulls, $g$ groups in total, in such a way that within each group, each of its members has the same amount of money and the same amount of stocks. In this way there are given initial amounts of investors $N_1(0),N_2(0),\ldots,N_g(0)\in\mathbb{N}$ of each group, and when choosing an investor of group $i$, the price multiplies itself by a positive number $f_i$, and said investor leaves the group to become an inactive investor. The coefficients $0<f_1<f_2<\ldots<f_g$ represent that the investors of group $1$ are the bears with the greatest amount of stocks (so when they invest the price falls the most), while the investors of group $g$ are the bulls with the greatest amount of money (so when they invest the price rises the most). A model without bears can also be represented: group $1$ would be the bulls with the tiniest amount of money, so when they invest the price raises the least. Analogously, a model without bulls can be represented.\newline

In this model the formula for the moments of the logarithmic return at time $t$ is
\begin{align}
\mathbb{E}\left(\log\left(\frac{p(t)}{p(0)}\right)^n\right)=\sum_{\substack{n_{1},\ldots,n_{g}\in\mathbb{N}_0\\n_{1}+\ldots+n_{g}=n}}\frac{n!}{\prod_{h}n_{h}!}\prod_{h=1}^g(\log f_h)^{n_{h}}\sum_{k_1=0}^{n_{1}}\frac{N_1(0)!}{(N_1(0)-k_1)!}\left\{\begin{matrix}n_{1}\\ k_1\end{matrix}\right\}\cdots\nonumber\\
\sum_{k_g=0}^{n_{g}}\frac{N_g(0)!}{(N_g(0)-k_g)!}\left\{\begin{matrix}n_{g}\\ k_g\end{matrix}\right\}\sum_{j=0}^{k_1+\ldots+k_g}(-1)^j\binom{k_1+\ldots+k_g}{j}\left(1-\frac{j}{N}\right)^{t}\text{ .}\label{form_2g}
\end{align}
The proof is analogous to the one in section \ref{momentos}.

\subsection{Fitting moments}

In this section, we generalize the procedure described in section \ref{fit_mom}.\newline

Assume that we are given the first $2g$ sample moments of the logarithmic returns: $m_n$ for $1\leq n\leq 2g$. Let
\begin{align*}
    v_1(x)&:=1+\sum_{k=1}^{2g}\frac{m_k}{k!}x^k\text{ ,}\\
    u_1(x)&:=\log v_1(x)\qquad\text{and}\\
    \kappa_n&:=u_1^{(n)}(0)\text{ .}
\end{align*}
Let $q(x)$ be the polynomial
\begin{align*}
    q(x):=\det\left(\begin{pmatrix}
        \kappa_1&\kappa_2&\cdots&\kappa_g\\
        \kappa_2&\kappa_3&\cdots&\kappa_{g+1}\\
        \vdots&\vdots&&\vdots\\
        \kappa_g&\kappa_{g+1}&\cdots&\kappa_{2g-1}
    \end{pmatrix}x-\begin{pmatrix}
        \kappa_2&\kappa_3&\cdots&\kappa_{g+1}\\
        \kappa_3&\kappa_4&\cdots&\kappa_{g+2}\\
        \vdots&\vdots&&\vdots\\
        \kappa_{g+1}&\kappa_{g+2}&\cdots&\kappa_{2g}
    \end{pmatrix}\right)\text{ ,}
\end{align*}
and let $\{r_h\}_{h=1}^g$ be its roots. Let
\begin{align*}
    V&:=\begin{pmatrix}
        1&r_1&\cdots &r_1^{g-1}\\
        \vdots &\vdots && \vdots\\
        1&r_g &\cdots &r_g^{g-1}
    \end{pmatrix}\text{ ,}\\
    D&:=\begin{pmatrix}
        r_1&&\\
        &\ddots &\\
        &&r_g
    \end{pmatrix}\qquad\text{and}\\
    \lambda&:=D^{-1}(V^T)^{-1}\begin{pmatrix}
        \kappa_1\\
        \vdots\\
        \kappa_g
    \end{pmatrix}\text{ .}\label{lambda_2}
\end{align*}

Under the following conditions, we can approximate the given sample moments:
\begin{itemize}
    \item $r_h\in\mathbb{R}\backslash\{0\}$ for $1\leq h\leq g$,
    \item $r_i\neq r_j$ for $i\neq j$ and
    \item $\lambda_h>0$ for $1\leq h\leq g$.
\end{itemize}
Assume these conditions hold, the given moments can be approximated as follows. Assume w.l.o.g. that $r_1<\ldots<r_g$.

Let us say we want to approximate the first $2g$ moments with an error less than some given $\varepsilon>0$.
\begin{procedure}
    \begin{enumerate}
        \item Set $C\gg 0$.\label{step_Nu_multi}
        \item Set $N_h(0)=\lfloor\lambda_h C\rfloor$ and $f_h=\exp r_h$.
        \item Compute the first $2g$ moments using the expression
        \begin{align*}
            \sum_{\substack{n_{1},\ldots,n_{g}\in\mathbb{N}_0\\n_{1}+\ldots+n_{g}=n}}\frac{n!}{\prod_{h}n_{h}!}\prod_{h=1}^g r_h^{n_{h}}\sum_{k_1=0}^{n_{1}}\frac{N_1(0)!}{(N_1(0)-k_1)!}\left\{\begin{matrix}n_{1}\\ k_1\end{matrix}\right\}\cdots\sum_{k_g=0}^{n_{g}}\frac{N_g(0)!}{(N_g(0)-k_g)!}\left\{\begin{matrix}n_{g}\\ k_g\end{matrix}\right\}C^{-\sum_h k_h}\text{ .}
        \end{align*}
        If the computed moments don't approximate the sample moments with an error less than $\varepsilon$, go back to step \ref{step_Nu_multi} and take a larger $C$.
        \item Set $N\geq\sum_{h=1}^g N_h(0)$.\label{step_N_multi}
        \item Set $t=\lfloor-N\log\left(1-C^{-1}\right)\rfloor$.
        \item Compute the first $2g$ moments using equation \eqref{form_2g}. If the computed moments don't approximate the sample moments with an error less than $\varepsilon$, go back to step \ref{step_N_multi} and take a larger $N$.
    \end{enumerate}
\end{procedure}
The proof of correctness of this procedure is analogous to the one in section \ref{fit_mom}.

\section{Conclusions}

In this work, we developed a model similar to the classical binomial one, but without the possibility of taking short positions. We think that this is relevant because there are many financial markets where short selling is restricted in one way or another.\newline
The model was described as a representation of a real scenario. In such scenario there are two groups of investors, the bulls and the bears. The bulls buy stocks, while the bears sell them. Mediating the transactions between the groups, there is a market maker that fixes the prices.

A formalization is provided in the appendix: a probability space was built that allows to represent the events and the information of the model.

We proved a formula that allows the computation of every moment of the logarithmic return. We also deduced a formula for the case with infinitely many investors. An important application of this last formula is that it allowed us to compute the moments of the classical binomial model.

As an application of the model, we presented a procedure for choosing the parameters, in order to approximate the first four sample moments arbitrarily close. This is relevant because the data show that the logarithmic returns deviate from normality\cite{cont2001empirical,Cont2007,mandelbrot1967variation,pagan1996econometrics,rachev2005empirical}. Further evidence of that was provided in the following section, where we also applied the procedure to real financial data and discovered a clear convergence pattern. This pattern, however, still requires an explanation.

We generalized the model for the case in which there are several groups of bulls and/or several groups of bears. We also generalized the algorithm given in section \ref{fit_mom}, which allows us to approximate $2g$ sample moments. These generalizations provide a framework for the integration of more complex phenomena.

\bibliographystyle{plain}
\bibliography{main}

\begin{thebibliography}{10}

\bibitem{akhiezer2020classical}
Naum~Ilyich Akhiezer.
\newblock {\em The classical moment problem and some related questions in
  analysis}.
\newblock Oliver \& Boyd, 1965.

\bibitem{caginalp2003theoretical}
Gunduz Caginalp and Donald Balenovich.
\newblock A theoretical foundation for technical analysis.
\newblock {\em Journal of Technical Analysis}, 59:5--21, 2003.

\bibitem{cont2001empirical}
Rama Cont.
\newblock Empirical properties of asset returns: stylized facts and statistical
  issues.
\newblock {\em Quantitative finance}, 1(2):223, 2001.

\bibitem{Cont2007}
Rama Cont.
\newblock {\em Volatility Clustering in Financial Markets: Empirical Facts and
  Agent-Based Models}, pages 289--309.
\newblock Springer Berlin Heidelberg, Berlin, Heidelberg, 2007.

\bibitem{cont2000herd}
Rama Cont and Jean-Philipe Bouchaud.
\newblock Herd behavior and aggregate fluctuations in financial markets.
\newblock {\em Macroeconomic dynamics}, 4(2):170--196, 2000.

\bibitem{cox1979option}
John~C Cox, Stephen~A Ross, and Mark Rubinstein.
\newblock Option pricing: A simplified approach.
\newblock {\em Journal of Financial Economics}, 7(3):229--263, 1979.

\bibitem{feller1991introduction}
William Feller.
\newblock {\em An introduction to probability theory and its applications,
  Volume 2}, volume~2.
\newblock John Wiley \& Sons, 1991.

\bibitem{patashnik1994concrete}
Ronald~L Graham, Donald~E Knuth, and Oren Patashnik.
\newblock {\em Concrete Mathematics: A Foundation for Computer Science}.
\newblock Addison-Wesley, 2 edition, 1994.

\bibitem{harrison1979martingales}
J~Michael Harrison and David~M Kreps.
\newblock Martingales and arbitrage in multiperiod securities markets.
\newblock {\em Journal of Economic theory}, 20(3):381--408, 1979.

\bibitem{harrison1981martingales}
J~Michael Harrison and Stanley~R Pliska.
\newblock Martingales and stochastic integrals in the theory of continuous
  trading.
\newblock {\em Stochastic processes and their applications}, 11(3):215--260,
  1981.

\bibitem{klenke2013probability}
Achim Klenke.
\newblock {\em Probability theory: a comprehensive course}.
\newblock Universitext. Springer Science \& Business Media, 2013.

\bibitem{krein1977markov}
Mark~Grigorevich Krein and Adol'f~Abramovich Nudel'man.
\newblock {\em The Markov moment problem and extremal problems: ideas and
  problems of PL Cebysev and AA Markov and their further development},
  volume~50 of {\em Translations of Mathematical Monographs}.
\newblock American Mathematical Society, 1977.

\bibitem{mandelbrot1967variation}
Benoit Mandelbrot.
\newblock The variation of some other speculative prices.
\newblock {\em The Journal of Business}, 40(4):393--413, 1967.

\bibitem{pagan1996econometrics}
Adrian Pagan.
\newblock The econometrics of financial markets.
\newblock {\em Journal of empirical finance}, 3(1):15--102, 1996.

\bibitem{rachev2005empirical}
Svetlozar~T Rachev, Stoyan~V Stoyanov, Almira Biglova, and Frank~J Fabozzi.
\newblock An empirical examination of daily stock return distributions for us
  stocks.
\newblock In {\em Data analysis and decision support}, pages 269--281.
  Springer, 2005.

\bibitem{rendleman1979two}
Richard~J Rendleman{, Jr.} and Britt~J Bartter.
\newblock Two-state option pricing.
\newblock {\em The Journal of Finance}, 34(5):1093--1110, 1979.

\bibitem{riordan1937moment}
John Riordan.
\newblock Moment recurrence relations for binomial, poisson and hypergeometric
  frequency distributions.
\newblock {\em The Annals of Mathematical Statistics}, 8(2):103--111, 1937.

\bibitem{schmudgen2017moment}
Konrad Schm{\"u}dgen.
\newblock {\em The moment problem}.
\newblock Graduate Texts in Mathematics. Springer International Publishing,
  2017.

\bibitem{sharpe1999investment}
William~F Sharpe, Gordon~J Alexander, and Jeffery~V Bailey.
\newblock {\em Investments}.
\newblock Prentice Hall Incorporated, 1999.

\bibitem{shohat1950problem}
James~Alexander Shohat and Jacob~David Tamarkin.
\newblock {\em The Problem of Moments}, volume~1 of {\em Mathematical Surveys
  and Monographs}.
\newblock American Mathematical Society, 1943.

\end{thebibliography}

\appendix

\section{Construction of the probability space}\label{proba}

We formalize the model presented in section \ref{palabras} with the following probability space. We consider the sample space
\begin{align*}
\Omega=\{U,D,I\}^{\mathbb{N}_0}\text{ ,}
\end{align*}
whose elements $\omega\in\Omega$ represent states of the world\cite{harrison1979martingales,harrison1981martingales}. For each $l\in\mathbb{N}_0$ and for each $\hat{\omega}\in\{U,D,I\}^l$, we have the set
\begin{align*}
A_{\hat{\omega}}=\{\omega\in\Omega:(\omega_0,\ldots,\omega_{l-1})=\hat{\omega}\}\text{ .}
\end{align*}
Let $\mathcal{F}_t$ be the $\sigma$-algebra generated by the family
\begin{align*}
\{A_{\hat{\omega}}\}_{\hat{\omega}\in\{U,D,I\}^{t}}\text{ ,}
\end{align*}
and
\begin{align*}
\mathcal{F}_{\infty}:=\bigcup_{t=0}^{\infty}\mathcal{F}_t\text{ .}
\end{align*}
Consider $q:\{(n_u,n_d)\in\mathbb{N}_0:n_u\leq N_u(0),n_d\leq N_d(0)\}\times\{U,D,I\}\to [0,1]$ given by
\begin{align*}
q(n_u,n_d,U):=\frac{N_u(0)-n_u}{N}\text{ ,}\qquad q(n_u,n_d,D):=\frac{N_d(0)-n_d}{N}\\
\text{and}\qquad q(n_u,n_d,I):=\frac{N-N_u(0)-N_d(0)+n_u+n_d}{N}\text{ ,}
\end{align*}
and $\mathbb{P}_t:\{A_{\hat{\omega}}\}_{\hat{\omega}\in\{U,D,I\}^{t}}\to\mathbb{R}$ recursively defined by
\begin{align*}
\mathbb{P}_t(A_{\hat{\omega}}):=\mathbb{P}_{t-1}(A_{(\hat{\omega}_0,\ldots,\hat{\omega}_{t-2})})\cdot q\left(\#\{i\leq t-2:\hat{\omega}_i=U\},\#\{i\leq t-2:\hat{\omega}_i=D\},\hat{\omega}_{t-1}\right)\\
\text{and}\qquad\mathbb{P}_0(\Omega):=1\text{ .}
\end{align*}

\begin{proposition}
$\mathbb{P}_t$ takes nonnegative values and
\begin{align*}
\sum_{\hat{\omega}\in\{U,D,I\}^{t}}\mathbb{P}_t(A_{\hat{\omega}})=1\,\text{.}
\end{align*}
\end{proposition}
\begin{proof}
We give a proof by induction on $t$. If $t=0$, it holds trivially. Assume that it holds until $t-1$.

If $P_t(A_{\hat{\omega}})<0$, then $\mathbb{P}_{t-1}(A_{(\hat{\omega}_0,\ldots,\hat{\omega}_{t-2})})>0$ and\newline
$q\left(\#\{i\leq t-2:\hat{\omega}_i=U\},\#\{i\leq t-2:\hat{\omega}_i=D\},\hat{\omega}_{t-1}\right)<0$.

If $\hat{\omega}_{t-1}=U$, then $\#\{i\leq t-2:\hat{\omega}_i=U\}>N_u(0)$. Let $t'$ be the greatest $i\leq t-2$ such that $\hat{\omega}_i=U$, then $\#\{i\leq t'-1:\hat{\omega}_i=U\}\geq N_u(0)$, but
\begin{align*}
0\leq\mathbb{P}_{t'+1}(A_{(\hat{\omega}_0,\ldots,\hat{\omega}_{t'})})\\
=\mathbb{P}_{t'}(A_{(\hat{\omega}_0,\ldots,\hat{\omega}_{t'-1})})\cdot q\left(\#\{i\leq t'-1:\hat{\omega}_i=U\},\#\{i\leq t'-1:\hat{\omega}_i=D\},U\right)\leq 0\text{ ,}
\end{align*}
hence $\mathbb{P}_{t'+1}(A_{(\hat{\omega}_0,\ldots,\hat{\omega}_{t'})})=0$. Consequently, $\mathbb{P}_{t-1}(A_{(\hat{\omega}_0,\ldots,\hat{\omega}_{t-2})})=0$, which is a contradiction. Analogously, one reaches a contradiction if $\hat{\omega}_{t-1}=D$. It cannot be that $\hat{\omega}_{t-1}=I$, because $N\geq N_u(0)+N_d(0)$, so $N+\#\{i\leq t-2:\hat{\omega}_i=U\}+\#\{i\leq t-2:\hat{\omega}_i=D\}\geq N_u(0)+N_d(0)$ and $q\left(\#\{i\leq t-2:\hat{\omega}_i=U\},\#\{i\leq t-2:\hat{\omega}_i=D\},I\right)\geq 0$.

That
\begin{align*}
    \sum_{\hat{\omega}\in\{U,D,I\}^{t}}\mathbb{P}_t(A_{\hat{\omega}})=1
\end{align*}
is easily seen using the recursive definition of $\mathbb{P}_t$.
\end{proof}

Then, $\mathbb{P}_t$ extends to a probability measure defined on $\mathcal{F}_t$. Observe that if $\hat{\omega}\in\{U,D,I\}^t$, then

\begin{equation}
\mathbb{P}_{s}(A_{\hat{\omega}})=\mathbb{P}_{s-1}(A_{\hat{\omega}})\label{colimit}
\end{equation}

for all $s>t$, and therefore $\mathbb{P}_s|_{\mathcal{F}_t}=\mathbb{P}_t$. Therefore, consider $\mathbb{P}_{\infty}:\mathcal{F}_{\infty}\to\mathbb{R}$, if $A\in\mathcal{F}_t$,
\begin{align*}
\mathbb{P}_{\infty}(A):=\mathbb{P}_{t}(A)\text{ .}
\end{align*}
This is well-defined because of \eqref{colimit}. $\mathcal{F}_{\infty}$ is an algebra of sets and $\mathbb{P}_{\infty}:\mathcal{F}_{\infty}\to [0,1]$ is finitely additive.

\begin{proposition}
Let $\{A_n\}_{n\in\mathbb{N}}\subset\mathcal{F}_{\infty}$ be a pairwise disjoint family of sets such that $A=\bigcup_{n\in\mathbb{N}}A_n\in\mathcal{F}_{\infty}$. Then
\begin{align*}
    \mathbb{P}_{\infty}(A)=\sum_{n=1}^{\infty}\mathbb{P}_{\infty}(A_n)\,\text{.}
\end{align*}
\end{proposition}
\begin{proof}
It's easy to prove that
\begin{align*}
    \sum_{n=1}^{\infty}\mathbb{P}_{\infty}(A_n)\leq\mathbb{P}_{\infty}(A)\text{ .}
\end{align*}

To verify that
\begin{align*}
    \sum_{n=1}^{\infty}\mathbb{P}_{\infty}(A_n)\geq\mathbb{P}_{\infty}(A)\,\text{,}
\end{align*}
it suffices to prove that it exists $N\in\mathbb{N}$ such that
\begin{align*}
    A\subset\bigcup_{n=1}^N A_n\,\text{,}
\end{align*}
and a proof of this can be found in example 1.63 of \cite{klenke2013probability}.
\end{proof}

Using the Hahn-Kolmogorov theorem, one can then extend $\mathbb{P}_{\infty}$ to a unique probability measure $\mathbb{P}:\mathcal{F}\to [0,1]$, where $\mathcal{F}$ is the $\sigma$-algebra generated by $\mathcal{F}_{\infty}$.

\section{Proofs}

\subsection{Computation of moments of the logarithmic returns}\label{app_mom_tri}

\begin{proof}[Proof of lemma \ref{lemma_first_formula}]
We give a proof by induction on $k$. For $k=0$, the left-hand side is
\begin{align*}
\mathbb{E}\left(\frac{N_u(i_{1}-1)!}{(N_u(i_{1}-1)-l_u)!}\frac{N_d(i_{1}-1)!}{(N_d(i_{1}-1)-l_d)!}\right)\text{ .}
\end{align*}
Observe that
\begin{align*}
\mathbb{E}\left(\left.\frac{N_u(i+1)!}{(N_u(i+1)-l_u)!}\frac{N_d(i+1)!}{(N_d(i+1)-l_d)!}\right|\mathcal{F}_{i}\right)=\frac{N_u(i)!}{(N_u(i)-l_u)!}\frac{N_d(i)!}{(N_d(i)-l_d)!}\left(1-\frac{l_u+l_d}{N}\right)\text{ .}
\end{align*}
By successive use of the tower rule we get
\begin{align*}
\mathbb{E}\left(\frac{N_u(i_{1}-1)!}{(N_u(i_{1}-1)-l_u)!}\frac{N_d(i_{1}-1)!}{(N_d(i_{1}-1)-l_d)!}\right)=\frac{N_u(0)!}{(N_u(0)-l_u)!}\frac{N_d(0)!}{(N_d(0)-l_d)!}\left(1-\frac{l_u+l_d}{N}\right)^{i_1-1}\text{ ,}
\end{align*}
which proves the case $k=0$.

Assume that \eqref{prop_3} holds for $k$. Successively using the tower rule, it's straightforward to see that \eqref{prop_3} holds for $k+1$.
\end{proof}

\begin{proof}[Proof of lemma \ref{coro_mom_tri}]
Let
\begin{align*}
B_n:=\left\{(k,\overrightarrow{n},s):1\leq k\leq n,\,\overrightarrow{n}\in\mathbb{N}^k\text{ such that }\left|\overrightarrow{n}\right|_1=n,\, s\in\{u,d\}^k\right\}\text{ ,}
\end{align*}
then we get
\begin{align*}
    \mathbb{E}\left(\log\left(\frac{p(t)}{p(0)}\right)^n\right)=\sum_{(k,\overrightarrow{n},s)\in B_n}\sum_{\substack{m_i\in\mathbb{N}\\m_1+\ldots+m_k\leq t}}\frac{n!}{n_1!\cdots n_k!}\frac{1}{N^k}\\
\cdot\prod_{j=1}^k\left(1-\frac{k-j+1}{N}\right)^{m_j-1}\frac{N_u(0)!}{(N_u(0)-\#\{s_j=u\})!}\frac{N_d(0)!}{(N_d(0)-\#\{s_j=d\})!}\prod_{j=1}^k(\log s_j)^{n_j}\text{ .}
\end{align*}
Let
\begin{align*}
C_n:=\left\{(n_u,n_d,\overrightarrow{n_u},\overrightarrow{n_d}):n_u,n_d\in\mathbb{N}_0\text{ such that }n_u+n_d=n,\,\overrightarrow{n_u}\in\bigcup_{k_u\geq 0}\mathbb{N}^{k_u}\text{ such that }\left|\overrightarrow{n_u}\right|_1=n_u,\right.\\
\left.\overrightarrow{n_d}\in\bigcup_{k_d\geq 0}\mathbb{N}^{k_d}\text{ such that }\left|\overrightarrow{n_d}\right|_1=n_d\right\}\text{ .}
\end{align*}
There is a surjective assignment from $B_n$ to $C_n$, where
\begin{align*}
n_u:=\sum_{j:s_j=u}n_j\text{ ,}
\end{align*}
$\overrightarrow{n_u}$ is the vector $\overrightarrow{n}$ with only the components $n_j$ such that $s_j=u$, and giving analogous definitions for $n_d$ and $\overrightarrow{n_d}$. The points in the preimage of $(n_u,n_d,\overrightarrow{n_u},\overrightarrow{n_d})$ are associated with terms of the form
\begin{align*}
\sum_{\substack{m_i\in\mathbb{N}\\m_1+\ldots+m_{k_u+k_d}\leq t}}\frac{n!}{n_1!\cdots n_{k_u+k_d}!}\frac{1}{N^{k_u+k_d}}\\
\cdot\prod_{j=1}^{k_u+k_d}\left(1-\frac{k_u+k_d-j+1}{N}\right)^{m_j-1}\frac{N_u(0)!}{(N_u(0)-k_u)!}\frac{N_d(0)!}{(N_d(0)-k_d)!}(\log u)^{n_u}(\log d)^{n_d}\text{ .}
\end{align*}
The preimage of $(n_u,n_d,\overrightarrow{n_u},\overrightarrow{n_d})$ has $(k_u+k_d)!/(k_u!k_d!)$ elements, therefore
\begin{align}
\mathbb{E}\left(\log\left(\frac{p(t)}{p(0)}\right)^n\right)=\sum_{\substack{n_u,n_d\in\mathbb{N}_0\\n_u+n_d=n}}\sum_{\substack{n_u^i\in\mathbb{N}\\n_u^1+\ldots+n_u^{k_u}=n_u}}\sum_{\substack{n_d^i\in\mathbb{N}\\n_d^1+\ldots+n_d^{k_d}=n_d}}\frac{(k_u+k_d)!}{k_u!k_d!}\nonumber\\
\cdot\sum_{\substack{m_i\in\mathbb{N}\\m_1+\ldots+m_{k_u+k_d}\leq t}}\frac{n!}{n_u^1!\cdots n_u^{k_u}!\cdot n_d^1!\cdots n_d^{k_d}!}\frac{1}{N^{k_u+k_d}}\prod_{j=1}^{k_u+k_d}\left(1-\frac{k_u+k_d-j+1}{N}\right)^{m_j-1}\nonumber\\
\cdot\frac{N_u(0)!}{(N_u(0)-k_u)!}\frac{N_d(0)!}{(N_d(0)-k_d)!}(\log u)^{n_u}(\log d)^{n_d}\nonumber\\
=\sum_{\substack{n_u,n_d\in\mathbb{N}_0\\n_u+n_d=n}}(\log u)^{n_u}(\log d)^{n_d}\sum_{\substack{n_u^i\in\mathbb{N}\\n_u^1+\ldots+n_u^{k_u}=n_u}}\frac{N_u(0)!}{N^{k_u}(N_u(0)-k_u)!}\sum_{\substack{n_d^i\in\mathbb{N}\\n_d^1+\ldots+n_d^{k_d}=n_d}}\frac{N_d(0)!}{N^{k_d}(N_d(0)-k_d)!}\nonumber\\
\cdot\frac{(k_u+k_d)!}{k_u!k_d!}\frac{n!}{n_u^1!\cdots n_u^{k_u}!\cdot n_d^1!\cdots n_d^{k_d}!}\sum_{\substack{m_i\in\mathbb{N}_0\\m_0+\ldots+m_{k_u+k_d}=t-k_u-k_d}}\prod_{j=0}^{k_u+k_d}\left(1-\frac{j}{N}\right)^{m_j}\text{ .}\tag{\ref{mom1}}
\end{align}
\end{proof}

\begin{proof}[Proof of lemma \ref{comb_lemma}]
By induction on $k$. If $k=0$, it obviously holds. Assume that it holds for $k$, then
\begin{align*}
\sum_{m_0+\ldots+m_k+m_{k+1}=m}\prod_{j=0}^{k+1}c_j^{m_j}=\sum_{l=0}^m\sum_{m_0+\ldots+m_{k-1}=m-l}\sum_{m_{k}=0}^lc_k^{m_k}c_{k+1}^{l-m_k}\prod_{j=0}^{k-1}c_j^{m_j}\\
=\sum_{l=0}^m\frac{c_{k+1}^{l+1}-c_{k}^{l+1}}{c_{k+1}-c_{k}}\sum_{m_0+\ldots+m_{k-1}=m-l}\prod_{j=0}^{k-1}c_j^{m_j}\\
=\frac{c_{k+1}}{c_{k+1}-c_{k}}\sum_{m_0+\ldots+m_{k-1}+l=m}c_{k+1}^{l}\prod_{j=0}^{k-1}c_j^{m_j}-\frac{c_{k}}{c_{k+1}-c_{k}}\sum_{m_0+\ldots+m_{k-1}+l=m}c_{k}^{l}\prod_{j=0}^{k-1}c_j^{m_j}\text{ .}
\end{align*}
Using the inductive hypothesis here and manipulating the resulting expression, we get that it holds for $k+1$.
\end{proof}

\subsection{Fitting moments}\label{app_fit_mom}

\begin{proof}[Proof of proposition \ref{prop_mom}]
The moment-generating function of $Y$ is
\begin{align*}
    \mathbb{E}\left(\exp(sY)\right)=\exp\left(\lambda_1\left(\exp(r_1 s)-1\right)+\lambda_2\left(\exp(r_2 s)-1\right)\right)\text{ .}
\end{align*}
    Let
\begin{align*}
    u_2(s)&:=\lambda_1\left(\exp(r_1 s)-1\right)+\lambda_2\left(\exp(r_2 s)-1\right)\qquad\text{and}\\
    v_2(s)&:=\exp(u_2(s))\text{ ,}
\end{align*}
then we want to prove that
\begin{align*}
    v_2^{(n)}(0)=v_1^{(n)}(0)
\end{align*}
for $1\leq n\leq 4$.\newline
It can be shown by induction that
\begin{align*}
    u_i^{(n)}(s)=v_i^{(n)}(s)v_i(s)^{-1}+P_n(v_i'(s),\ldots,v_i^{(n-1)}(s),v_i(s)^{-2},\ldots,v_i(s)^{-n})\text{ ,}
\end{align*}
where $P_n$ is a polynomial in several variables, and using this, we reduce the problem to proving that
\begin{align*}
    u_2^{(n)}(0)=u_1^{(n)}(0)\text{ .}
\end{align*}
On the left-hand side, observe that
\begin{align*}
    u_2^{(n)}(s)&=\lambda_1 r_1^n\exp(r_1 s)+\lambda_2 r_2^n\exp(r_2 s)\qquad\text{and}\\
    u_2^{(n)}(0)&=\lambda_1 r_1^n+\lambda_2 r_2^n\text{ ,}
\end{align*}
so
\begin{align*}
    \begin{pmatrix}
        u_2^{(1)}(0)\\
        \vdots\\
        u_2^{(4)}(0)
    \end{pmatrix}=\begin{pmatrix}
        V^T\\
        V^TD^2
    \end{pmatrix}D\lambda=\begin{pmatrix}
        V^T\\
        V^TD^2
    \end{pmatrix}(V^T)^{-1}\begin{pmatrix}
        \kappa_1\\
        \kappa_2
    \end{pmatrix}\text{ .}
\end{align*}
On the two last rows of this computation, observe that
\begin{align*}
    V^TD^2(V^T)^{-1}=\left(V^TD(V^T)^{-1}\right)^2\text{ .}
\end{align*}
Let
\begin{equation}
    C:=\begin{pmatrix}
        \kappa_1&\kappa_2\\
        \kappa_2&\kappa_3
    \end{pmatrix}^{-1}\begin{pmatrix}
        \kappa_2&\kappa_3\\
        \kappa_3&\kappa_4
    \end{pmatrix}=\begin{pmatrix}
        0&-c_0\\
        1&-c_{1}
    \end{pmatrix}\text{ ,}\label{companion}
\end{equation}
then
\begin{align*}
    q(x)&=\det\begin{pmatrix}
        \kappa_1&\kappa_2\\
        \kappa_2&\kappa_3
    \end{pmatrix}\det(x-C)\qquad\text{and}\\
    \det(x-C)&=(x-r_1)(x-r_2)\text{ .}
\end{align*}
It's known and trivial that
\begin{align*}
    VC=DV\text{ ,}
\end{align*}
hence
\begin{align*}
    V^TD(V^T)^{-1}=C^T
\end{align*}
and
\begin{align*}
    \begin{pmatrix}
        u_2^{(1)}(0)\\
        \vdots\\
        u_2^{(4)}(0)
    \end{pmatrix}=\begin{pmatrix}
        I\\
        (C^T)^2
    \end{pmatrix}\begin{pmatrix}
        \kappa_1\\
        \kappa_2
    \end{pmatrix}\text{ .}
\end{align*}
Observe that
\begin{align*}
    C^T\begin{pmatrix}
        a_0\\
        \vdots\\
        a_{1}
    \end{pmatrix}=\begin{pmatrix}
        a_1\\
        -c_0 a_0-c_{1}a_{1}
    \end{pmatrix}\text{ .}
\end{align*}
Therefore
\begin{align*}
    (C^T)^2\begin{pmatrix}
        \kappa_1\\
        \kappa_2
    \end{pmatrix}
\end{align*}
is the vector given by
\begin{align*}
    \begin{pmatrix}
        a_{2}\\
        a_{3}
    \end{pmatrix}
\end{align*}
of the sequence recursively defined by
\begin{align*}
    a_{n+2}:=-c_0 a_{n}-c_1 a_{n+1}
\end{align*}
and starting at
\begin{align*}
    \begin{pmatrix}
        a_0\\
        a_{1}
    \end{pmatrix}:=\begin{pmatrix}
        \kappa_1\\
        \kappa_2
    \end{pmatrix}\text{ .}
\end{align*}
From the last column of equation \eqref{companion}, we get
\begin{align*}
    \begin{pmatrix}
        \kappa_{3}\\
        \kappa_{4}
    \end{pmatrix}=\begin{pmatrix}
        \kappa_1&\kappa_2\\
        \kappa_2&\kappa_3
    \end{pmatrix}\begin{pmatrix}
        -c_0\\
        -c_{1}
    \end{pmatrix}\text{ ,}
\end{align*}
and from this we conclude that
\begin{align*}
    (C^T)^2\begin{pmatrix}
        \kappa_1\\
        \kappa_2
    \end{pmatrix}=\begin{pmatrix}
        \kappa_{3}\\
        \kappa_{4}
    \end{pmatrix}\text{ ,}
\end{align*}
which completes the proof.
\end{proof}

\end{document}